\documentclass[11pt]{article}
\usepackage{a4wide,amsmath,amssymb,comment}
\usepackage[mathscr]{eucal}
\usepackage{jheppub}
\def\zi{\mathbb{Z}}

\newcommand{\oao}[2]{{#1\atopwithdelims[]#2}}

\title{Asymmetric Gepner models in type II}

\author[a,b]{Dan Isra\"el}
\author[a,b]{and Vincent Thi\'ery}
\affiliation[a]{
Sorbonne Universit\'es, UPMC Univ Paris 06, UMR 7589, LPTHE, F-75005, 
Paris, France
}
\affiliation[b]{
CNRS, UMR 7589, LPTHE, F-75005, Paris, France
}

\emailAdd{israel@lpthe.jussieu.fr}
\emailAdd{vjmthiery@gmail.com}

\abstract{We describe new four-dimensional type II compactifications with $\mathcal{N}=2$ supersymmetry, based on 
asymmetric Gepner models for $K3 \times T^2$. In more than half of these models, all the $K3$ moduli are lifted, giving 
at low energies $\mathcal{N}=2$ supergravity with the $STU$ vector multiplets and no hypermultiplets. 
}

\begin{document}
\maketitle

\section{Introduction}

Understanding compactifications with fewer moduli and fewer supersymmetries is still a major goal of string theory. Besides 
compactifications with Ramond-Ramond fluxes, that are quite successful in this respect but lack a usable worldsheet formulation, 
it is desirable to find models with a better grip on $\alpha'$ corrections beyond the supergravity regime.

Unlike in heterotic strings, it is not possible to consider type II compactifications with NSNS-fluxes only, since the three-form $H$ is 
closed. It leaves us the possibility of using non-geometric fluxes, described either as asymmetric orbifolds 
of rational tori~\cite{Narain:1986qm,Dabholkar:1998kv,Anastasopoulos:2009kj}, 
using free-fermion constructions~\cite{Kawai:1986ah,Antoniadis:1986rn} or as (generalized) T-duals of so-called T-folds 
that are locally geometric~\cite{Dabholkar:2002sy,Hull:2004in,Hellerman:2002ax} (see~\cite{Hohm:2013bwa} and references 
therein for a recent review). Studying such  (geometric or non-geometric) fluxes in {\it interacting} rather than 
free worldsheet conformal field theories would allow to understand how these twists can be defined in 
non-trivial backgrounds, beyond the twisted tori or free-fermions examples. 

A large class of supersymmetric compactifications on Calabi-Yau manifolds, in their stringy regime of negative K\"ahler moduli, can be 
described by non-trivial superconformal field theories constructed by Gepner using $\mathcal{N}=(2,2)$ 
minimal models as building blocks~\cite{Gepner:1987qi,Gepner:1987vz}.  The structure of Gepner model is very rigid, 
being tightly constrained by modular invariance of the one-loop partition function. Nevertheless some asymmetric heterotic Gepner 
models with$(2,0)$ superconformal symmetry have been considered in the past, corresponding to (the small volume limit of) 
compactifications with non-standard embedding gauge bundles~\cite{Schellekens:1989wx,Berglund:1995dv,Kreuzer:2009ha,GatoRivera:2010gv}. 
To our knowledge, analogous supersymmetric constructions in type IIA/IIB superstrings have not been considered yet.\footnote{Different 
constructions of non-supersymmetric asymmetric models were considered in~\cite{GatoRivera:2007yi,Kawai:2007nb}.}

In this work we describe a simple method to construct asymmetric Gepner models in type IIA/B superstrings, starting with 
an $\mathcal{N}=4$ compactification on $K3\times T^2$ where the $K3$ surface is described by a Gepner model; 
they can be thought as some sort of non-geometric $T^2$ fibration. These asymmetric 
Gepner models have several interesting and unusual properties. They provide $\mathcal{N}=2$ compactifications to 
four dimensions such that all the eight space-time supercharges come from the 
left, as in~\cite{Vafa:1995gm}, unlike compactifications on CY threefolds for which four supercharges comes from the left and 
four from the right. Since the right Ramond ground states are massive, there are also no massless states from the whole RR sector. 

Having studied the massless spectra for all the 62 asymmetric models of this sort, we have found that they all have 
at most few remaining massless scalars in the spectrum, and that 33 of them are actually devoid 
of any remaining massless modulus from the $K3$ two-fold. Their field content corresponds at low energies to four-dimensional 
$\mathcal{N}=2$ supergravity with the $STU$ vector multiplets and no hypermultiplets. Some supersymmetric compactifications with few 
moduli were constructed using free-fermionic constructions~\cite{Ferrara:1989nm,Dolivet:2007sz} or freely acting toroidal 
orbifolds~\cite{Dabholkar:1998kv,Anastasopoulos:2009kj}. Our models provide a broad generalization of these models  
to interacting superconformal field theory compactifications.\footnote{As Gepner models with small levels 
for all minimal models are free actually theories the aforementioned examples should be given by specific cases of our construction.}

This note is organized as follows. In section~\ref{sec:review} we review the construction of $K3$ Gepner models. 
In section~\ref{sec:asy} we explain how to build 
the asymmetric Gepner models, provide their partition function and give some of their generic properties. 
In section~\ref{sec:spectra} we give an overview of the massless spectra for all models. 
Finally we summarize our findings and give some future directions of research in section~\ref{sec:concl}. We give a detailed 
list of all massless spectra in appendix~\ref{appspectra} and recall some basic facts about $\mathcal{N}=2$ characters in appendix~\ref{appchar}.

\section{Gepner models for K3}
\label{sec:review}
In this section we review the construction of Gepner models with a $K3$ target-space, using a slightly different method compared 
to the original work of Gepner, that is more convenient for our purposes. We also explain how to derive  their massless spectra.

Our construction starts with $K3\times T^2$ compactifications in type IIA/B superstrings, preserving $\mathcal{N}=4$ supersymmetry 
in four dimensions, where the $K3$ surface is chosen at a Gepner point in its moduli space~\cite{Gepner:1987vz}. 
Gepner models for $K3$, except for two out of the sixteen available, 
are made out of four $\mathcal{N}=(2,2)$ minimal models, whose {\it supersymmetric} levels will be denoted
 by $\{k_i,\, i=1,\ldots,4\}$. In these conventions (see app.~\ref{appchar}), the left and right central charges  
of a minimal model are equal to $3-6/k$. In particular a model with $k=2$ is trivial, 
being a SCFT with $(c,\bar c)=(0,0)$. 

The central charges of an $\mathcal{N}=(2,2)$ SCFT with a $K3$ target space should be $(c,\bar c)=(6,6)$, which 
translates into
\begin{equation}
\sum_{i=1}^4 \frac{1}{k_i}=1\, .
\end{equation} 
For simplicity of notation we consider only models with four minimal model factors in the following. These models are 
entirely specified by a quadruplet $(k_1,\ldots,k_4)$, where $k_i \geqslant 2$, if a diagonal modular-invariant is chosen 
for each affine $SU(2)$ spin as we shall assume for definiteness.

\subsection{Partition function}

The main achievement of Gepner's work is to find a modular-invariant combination of minimal models with a generalized GSO projection 
onto integer left and right R-charges, allowing to build a supersymmetric string compactification out of the model. 

In Gepner models, the GSO projection can be decomposed in two steps. The first step is a {\it diagonal} 
$\mathbb{Z}_K$ orbifold of the minimal models, with 
\begin{equation}
K=\text{lcm}\, (k_1,\ldots,k_4)\, ,
\end{equation}
giving integer left and right R-charges, while the second step involves two {\it chiral} 
$\mathbb{Z}_2$ orbifolds, enforcing that the former are both odd integers.

A type IIB modular invariant partition function for such a $K3 \times T^2$ compactification is given by the following expression:
\begin{multline}
\label{partuntwist}
Z  = \frac{\Gamma_{2,2} (T,U)}{\tau_2^2 \eta^4 \bar{\eta}^4} \frac{1}{K} \sum_{\gamma,\delta \in \mathbb{Z}_{K}} 
\frac{1}{2} \sum_{a,b=0}^{1} (-)^{a+b} \frac{1}{2} \sum_{\bar{a},\bar{b}=0}^{1} (-)^{\bar{a}+\bar{b} }
\frac{\vartheta^2 \oao{a}{b}}{\eta^2} \frac{\bar{\vartheta}^2 \oao{\bar{a}}{\bar{b}}}{\bar{\eta}^2}\\
\prod_{i=1}^{4}   \sum_{2j_i=0}^{k_i-2} \sum_{m_i \in \mathbb{Z}_{2k_i}} 
e^{i\pi (2\delta-b+\bar b)  \frac{m_i}{k_i}}
C^{j_i}_{m_i+a} \oao{a}{b} 
\bar{C}^{j_i}_{m_i+\bar{a}+2\gamma} \oao{\bar{a}}{\bar{b}} \, ,
\end{multline}
in terms of the minimal models characters $C^j_m \oao{a}{b}$, see app.~\ref{appchar}, and of the $T^2$ lattice $\Gamma_{2,2} (T,U)$. 
The sectors of the $\mathbb{Z}_{K}$ orbifold are labeled by $(\gamma,\delta)$, while those of the chiral $\mathbb{Z}_2$ 
projections are $(a,b)$ and $(\bar a, \bar b)$ respectively. An explicit check of modular invariance is 
presented in app.~\ref{appchar}.\footnote{We do not use the 'beta-method' introduced by Gepner in the original 
article~\cite{Gepner:1987qi}, as we find that our alternative formulation makes the computation of the massless spectrum 
easier for asymmetric models.}

\subsection{Massless spectrum}

Massless scalars in space-time are obtained by combining states of the left and right (anti-) chiral rings of the 
$\mathcal{N}=2$ superconformal algebra. A very detailed study of Gepner models spectra can be 
found $e.g.$ in~\cite{Fuchs:1989yv}; a more specific analysis of $K3$ Gepner models with an emphasis on 
the $(4,4)$ structure can be found in~\cite{Nahm:1999ps}.

In the case at hand, the $S$, $T$ and $U$ moduli of  $K3\times T^2$ compactifications 
are built using the identity operator in the Gepner model. Second, 
the $K3$ moduli are given by chiral/antichiral operators on the left and on the right with $|Q_R|=|\bar Q_R|=1$. 
In the case of (generic) CY threefolds, one has to consider four different rings, $(a,a)$, $(c,c)$, $(a,c)$ and $(a,a)$, depending 
on the choice of chiral or antichiral operators on both sides. In the case of K3 models, thanks to $\mathcal{N}=(4,4)$ 
superconformal symmetry, these four rings are related by (inner) automorphisms of $SU(2)_R\times \overline{SU(2)}_R$. 

\subsubsection*{Massless states in the (a,a) ring}
Since the four rings are isomorphic to each other, it is enough to consider one of them to get the full spectrum of 
massless scalars. We can choose for instance to study the $(a,a)$ spectrum.

Anti-chiral operators in each of the minimal models have even fermion number, $m=2j$ and conformal dimension\footnote{We have 
chosen the representation of chiral and antichiral states in relation with our choice of domain 
$m_i\in \{0,1,\ldots,2k_i-1\}$.} 
\begin{equation}
\Delta = - \frac{Q_{R,i}}{2} =\frac{m_i}{2k_i}= \frac{j_i}{k_i}\, , \ i=1,\ldots 4\, , 
\end{equation}
see appendix~\ref{appchar}. One gets then a left antichiral state with $\bar{Q}_R=-1$ provided that  
\begin{equation}
\label{masscondit}
\sum_{i=1}^4 \frac{m_i}{k_i}=\sum_{i=1}^4 \frac{2j_i}{k_i} = 1\, ,
\end{equation}
which, naturally, satisfies the $\mathbb{Z}_K$-orbifold invariance.  

On the right-moving side, once the $\{j_i\}$'s have been chosen in order to satisfy~(\ref{masscondit}), states in the untwisted sector 
($\gamma =0$) are automatically anti-chiral states with $\bar{Q}_R=-1$, hence giving at the end massless scalars in space-time.

In the twisted sectors ($\gamma \neq 0$), anti-chiral states  occur on the right 
if one of the two following conditions is satisfied {\it for every minimal model}:\footnote{Likewise, in the $(\star,c)$ rings 
one has either $\gamma \equiv 1 \mod k_i$ or $2j_i + \gamma  \equiv 0 \mod k_i$ for each minimal model.}
\begin{enumerate}
\item If $\gamma \equiv 0 \mod k_i$, due to the periodicity of minimal model characters, see eq.~(\ref{symMM}).
\item If $2j_i + \gamma +1 \equiv 0 \mod k_i$, the state equivalence  
$(j,2j+2\gamma,2)\sim (k/2-j-1,2j+2\gamma-k,0)$, see again eq.~\ref{symMM}, gives an anti-chiral state, as 
$2j+2\gamma-k_i\equiv k-2j-2\mod 2k_i$. However this map flips the GSO parity, see eq.~(\ref{reflsym}), 
hence it should be used in an {\it even} number of minimal models only.
\end{enumerate}
One easily checks that solutions of these constraints automatically give right anti-chiral states 
with $\bar{Q}_R=-1$, hence massless scalars again.

Once the massless states in the $(a,a)$ ring have been determined, massless states in the three other rings follow  
from $\mathcal{N}=(4,4)$ superconformal symmetry.

\subsubsection*{An example}
As an example we consider the $(3,3,4,12)$ Gepner model. Marginal operators in any of the four chiral rings satisfy 
the constraint 
\begin{equation}
4j_1 + 4j_2 + 3j_3 + j_4 = 6\, ,
\label{achiral}
\end{equation}
with the restrictions $0\leqslant 2j_{1,2} \leqslant 1$, $0\leqslant 2j_{3} \leqslant 2$ and $0\leqslant 2j_{4} \leqslant 10$. 

We use the notation $[2j_1,2j_2,2j_3,2j_4]_\gamma$, identifying massless states by their $SU(2)^4$ spins and their twisted 
sector $\gamma$.  One finds first the following ten $(a,a)$ operators in the untwisted sector:
\begin{equation}
\label{listchir}
\begin{array}{c}
\left[0,0,1,9\right]_0\\
\left[0, 1, 0, 8\right]_0 \, , \  \left[1, 0, 0, 8\right]_0  \, , \\
\left[0,0,2,6\right]_0 \\
\left[0, 1, 1, 5\right]_0 \, , \ \left[1, 0, 1, 5\right]_0 \, , \\
\left[1, 1, 0, 4\right]_0 \, , \\ 
\left[0, 1, 2, 2\right]_0 \, , \ \left[1, 0, 2, 2\right]_0 \, , \\
\left[1, 1, 1, 1\right]_0\, .
\end{array}
\end{equation}
For each of these operators, there is one and precisely one twisted sector such that the right 
operator is also anti-chiral and of right R-charge minus one:
\begin{equation}
\label{listchirtw}
\begin{array}{ccc}
\gamma = 2 & \longleftrightarrow & \left[0, 0, 1, 9\right]_2\\
\gamma = 3 & \longleftrightarrow & \left[0, 1, 0, 8\right]_3\, , \ \left[1, 0, 0, 8\right]_3\\
\gamma = 5 & \longleftrightarrow & \left[0, 0, 2, 6\right]_5\\
\gamma = 6 & \longleftrightarrow & \left[0, 1, 1, 5\right]_6 \, , \ \left[1, 0, 1, 5\right]_6 \\
\gamma = 7 & \longleftrightarrow & \left[1, 1, 0, 4\right]_7  \\
\gamma = 9 & \longleftrightarrow & \left[0, 1, 2, 2\right]_9\, , \ \left[1, 0, 2, 2\right]_9\\
\gamma = 10 & \longleftrightarrow & \left[1, 1, 1, 1\right]_{10}\\
\end{array}
\end{equation}
hence giving ten additional massless states in the $(a,a)$ ring. Notice that this one-to-one correspondence between 
the untwisted and twisted sector antichiral operators is accidental. It is not true for every $K3$ Gepner model. 

The same story holds for the $(c,c)$ and $(a,c)$ and $(c,a)$ rings. Altogether one finds $20+20+20+20=80$ operators. One has in 
addition the identity operator, that gives the universal $S$ modulus containing the dilaton and NSNS axion. One gets then, as expected, 
$81$ massless scalars spanning the moduli of $K3$ compactifications. On top of this, the two-torus provides the usual $T$ and 
$U$ moduli.

\section{Asymmetric Gepner models}
\label{sec:asy}
We now present a class of asymmetric Gepner models that provide new types of supersymmetric type II compactifications.\footnote{Lowest 
levels models at small radii may likely be similar  to some free-fermion constructions as~\cite{Dolivet:2007sz}.}

\subsection{The idea}
The basic idea behind these constructions is very simple.  Consider the combination of characters
\begin{equation}
\frac{\Theta_{m,k}}{\eta} \bar{\chi}^j \frac{\bar{\vartheta} \oao{\bar a}{\bar b}}{\bar \eta}\, ,
\label{charcomba}
\end{equation}
where $\chi^j$ is an affine $SU(2)$ character at level $k-2$ and $\Theta_{m,k}$ is a theta-function at level $k$, which gives the lattice of 
a holomorphic compact boson at radius $\sqrt{\alpha' k}$.  
	
The modular properties of~(\ref{charcomba}) are the same as those of the anti-holomorphic minimal model 
character $\bar{C}^{j}_{m} \oao{\bar a}{\bar b}$, thus we can trade the latter for the former in a Gepner 
model partition function without spoiling modular invariance. Crucially, as far as the $\mathbb{Z}_{2k}$ charge 
is concerned, an $\mathcal{N}=2$ minimal model character with $c=3-6/k$ transforms as the {\it conjugate} of a $U(1)$ character at 
level $k$.

A careful reader may be worried about right superconformal invariance, as one considers type II models which should 
have at least $\mathcal{N}=(1,1)$ superconformal symmetry. Let us consider the product of two terms like~(\ref{charcomba}), 
together with the corresponding left minimal model characters. The two right bosonic $SU(2)_{k_i-2}$ characters 
can be combined with free-fermion characters in order to make explicit an 
$\mathcal{N}=2$ superconformal symmetry on the right: 
\begin{multline}
\frac{\Theta_{m_3,k_3}}{\eta}\frac{\Theta_{m_4,k_4}}{\eta}  \frac{\vartheta \oao{a}{b}}{\eta}
\bar{\chi}^{j_3} \bar{\chi}^{j_4} \frac{\bar{\vartheta}^3 \oao{\bar a}{\bar b}}{\bar \eta^3}\\ = 
\sum_{\bar m_3 \in \mathbb{Z}_{2k_3}}\sum_{\bar m_4 \in \mathbb{Z}_{2k_4}}
\bar{C}^{j_3}_{\bar m_3} \oao{\bar a}{\bar b}  \bar{C}^{j_4}_{\bar m_4} \oao{\bar a}{\bar b} 
\left(\frac{\Theta_{m_3,k_3}}{\eta} \frac{\Theta_{m_4,k_4}}{\eta} \frac{\vartheta \oao{a}{b}}{\eta}\right)
\left(\frac{\bar{\Theta}_{\bar m_3,k_3}}{\bar \eta}
\frac{\bar{\Theta}_{\bar m_4,k_4}}{\bar \eta} \frac{\bar{\vartheta} \oao{\bar a}{\bar b}}{\bar \eta} \right)\, .
\label{charcombis}
\end{multline}
The decomposition~(\ref{charcombis}) shows first that this construction uses perfectly well-defined $\mathcal{N}=(2,2)$ 
superconformal field theories, as we get a product of ordinary holomorphic and anti-holomorphic characters for 
a couple of minimal models and for a couple of free $c=2$ theories.  Second, as the $\mathbb{Z}_{2k_i}$ charges 
of the minimal models are mixed with the lattice of the free bosons, it corresponds to 'fibering' $S^1$'s over minimal models. 
In order to get a four-dimensional theory at the end, one considers fibering at most a $T^2$ over the $K3$ Gepner model. 
For definiteness, one considers below the generic case with a non-degenerate $T^2$ fiber.

\subsection{Partition function}
Following the rules defined in the previous subsection, one obtains a modular-invariant partition function 
for an asymmetric Gepner model in type II by replacing in a $K3\times T^2$ ordinary Gepner model 
the right-moving characters of the last two minimal models and the $T^2$ contribution by a combination of characters 
of the type~(\ref{charcombis}). Explicitly, one gets:
\begin{multline}
\label{asympart}
Z = \frac{1}{\tau_2^2 \eta^2 \bar{\eta}^2} \frac{1}{K} \sum_{\gamma,\delta \in \mathbb{Z}_{K}} 
\frac{1}{2} \sum_{a,b=0}^{1} (-)^{a+b} \frac{1}{2} \sum_{\bar{a},\bar{b}=0}^{1} (-)^{\bar{a}+\bar{b} }
\frac{\vartheta^2 \oao{a}{b}}{\eta^2} \frac{\bar{\vartheta}^2 \oao{\bar{a}}{\bar{b}}}{\bar{\eta}^2}\\
\prod_{i=1}^{2}   \sum_{2j_i=0}^{k_i-2} \sum_{m_i \in \mathbb{Z}_{2k_i}} 
e^{i\pi (2\delta-b+\bar b)  \frac{m_i}{k_i}}
C^{j_i}_{m_i+a} \oao{a}{b} 
\bar{C}^{j_i}_{m_i+\bar{a}+2\gamma} \oao{\bar{a}}{\bar{b}} \\
\prod_{i=3}^{4}   \sum_{2j_i=0}^{k_i-2} \sum_{m_i,\bar{m}_i \in \mathbb{Z}_{2k_i}} 
e^{i\pi (2\delta-b+\bar b) \frac{m_i}{k_i}}
\left(\frac{\Theta_{m_i+\bar{a}+2\gamma,k_i}}{\eta}
\frac{\bar{\Theta}_{\bar m_i,k_i}}{\bar \eta} \right)
C^{j_i}_{m_i+a} \oao{a}{b} 
\bar{C}^{j_i}_{\bar m_i} \oao{\bar{a}}{\bar{b}} \, .
\end{multline}
This modular-invariant partition function corresponds to a well-defined type IIB string background, given that the underlying 
conformal field theory is unitary, has $(2,2)$ superconformal symmetry, left and right central charges 
$(c,\bar{c})=(12,12)$ and satisfies the requested spin-statistics connection in space-time thanks to the  GSO projection.

As we have noticed before, this model can  be thought as some sort of freely-acting 
asymmetric $\mathbb{Z}_{k_3} \times \mathbb{Z}_{k_4}$ orbifold of $K3\times T^2$ that acts as a shift in the lattice of the 
two-torus. This shift is asymmetric since the term in parenthesis in eq.~(\ref{asympart}) contains left and right 
$U(1)_{k_i}$ characters with different charges.

\subsection{Space-time supersymmetry and Ramond-Ramond ground states}

The left-moving sector of the SCFT defined by eq.~(\ref{asympart}) provides eight space-time supercharges, 
as the holomorphic part of its partition function  is the same as in an ordinary symmetric Gepner model, see eq.~(\ref{partuntwist}); 
it guarantees  a left $(c=6,\mathcal{N}=4)\times (c=3,\mathcal{N}=2)$ superconformal symmetry.

In contrast, there are no space-time supercharges coming from the right-moving sector. As follows from~(\ref{asympart}), 
the right R-charge of the $\mathcal{N}=2$ superconformal algebra is of the form (with $\bar{Q}_{fer}$ the right fermion number)
\begin{equation}
\bar{Q}_R =\bar{Q}_{fer} -\frac{m_1 +2\gamma}{k_1}-\frac{m_2 +2\gamma}{k_2} - \frac{\bar{m}_3}{k_3}- \frac{\bar{m}_4}{k_4}\mod 2\, ,
\end{equation}
where $\bar{m}_3$ and $\bar{m}_4$ are left unconstrained  by the GSO projection. This charge being generically fractional, it is 
not possible to achieve space-time supersymmetry using spectral flow of the right $\mathcal{N}=2$ superconformal 
algebra (see~\cite{Gepner:1989gr} for a detailed account on this mechanism).

To be more explicit, Ramond ground states in minimal models correspond to the quantum numbers $(j,m,a)=(0,\pm 1,1)$. In the 
right-moving sector, since $\bar{m}_3$ and $\bar{m}_4$ appear {\it both} as minimal model $\mathbb{Z}_{k_i}$ charges and in the $T^2$ lattice, 
two out of the four gravitini get a mass:\footnote{
The same holds for model with a single $S^1$ fiber, however there is only one term in the mass formula.}
\begin{equation}
\label{mass_shift}
M_{\psi_\mu} = \sqrt{\frac{1}{\alpha' k_3} +  \frac{1}{\alpha' k_4}}\, .
\end{equation}
Therefore, space-time four-dimensional $\mathcal{N}=4$ supersymmetry is broken to $\mathcal{N}=2$.  
Naturally the same holds for the RR ground states, 
which have the same mass shift~(\ref{mass_shift}) thanks to space-time supersymmetry.  

These $\mathcal{N}=2$ four-dimensional compactifications of type II superstrings are markedly distinct from usual 
compactifications on Calabi-Yau three-folds, as one has eight space-time supercharges from the left-movers, instead 
of having four supercharges from each side.

This demonstrates the non-geometric nature of these compactifications for the following reason.\footnote{Another 
more trivial indication of their non-geometric nature is simply that their partition function is asymmetric as we have already 
emphasized.} One can construct in principle 
non-linear sigma-models giving a different number of space-time supercharges from the left-movers and from the right-movers by 
adding H-flux (torsion) to the background, as the left- and right-handed worldsheet fermions couple to different torsionfull  
spin connections. It is not possible however to obtain type II compactifications with a standard supergravity limit 
having only NSNS-flux\footnote{In the non-compact case, no such obstruction exist, see $e.g.$~\cite{Maldacena:2000yy}.}. 
Typically non-geometric compactifications have a reduced number of moduli; we shall now study in detail the massless 
spectra of our models.

\section{Massless spectra of asymmetric models}
\label{sec:spectra}
The asymmetric Gepner models given by the partition function~(\ref{asympart}) have different numbers of 
massless moduli, depending on the values of $(k_1,\ldots,k_4)$, unlike the case of ordinary Gepner models 
for $K3$. We shall now study these massless spectra in detail, finding in particular that more than half of 
them have no massless moduli besides  $S$, $T$ and $U$.

\subsection{General rules}
On the left, the analysis of massless states starts similar to the case of ordinary minimal models that 
was considered in section~\ref{sec:review}, namely one has to look for elements of the chiral and  
antichiral rings with $|Q_R|=1$. 

There is however one important new ingredient. States in the $T^2$ theory have to be chiral or 
anti-chiral as well, which sets their left-moving momentum to zero. This gives a pair of non-trivial constraints, 
since the minimal model and $U(1)^2$ quantum numbers are mixed with each other, see~(\ref{asympart}). 
Explicitely one gets the  conditions
\begin{equation}
\label{asanti2}
 j_\ell + \gamma \equiv 0 \mod k_\ell \quad, \qquad \ell=3,4\, , 
\end{equation}
which already cut out a large part of the chiral and antichiral rings of the original Gepner model.

For each of the surviving left chiral/antichiral states one needs to check whether it is possible to obtain a massless state in space-time, 
$i.e.$ whether there exists a right primary state of dimension one-half associated with its quantum numbers. As the right R-charges are generically fractional rather than integer-valued, massless states are not necessarily associated with right chiral or antichiral states;  the usual argument relating massless states with BPS states of smallest R-charge in absolute value does not hold any more. 

In full generality, the right conformal dimension of conformal primaries that we need to consider, in the $\gamma$-th twisted sector, reads:
\begin{equation}
\bar \Delta = \bar{\Delta}_{1} +  \bar{\Delta}_{2} +
\frac{j_3(j_3+1)}{k_3} + \frac{j_4(j_4+1)}{k_4}
\end{equation}
where $\bar{\Delta}_\ell$ is the conformal dimension of the $\ell$th minimal model primary such that 
$m_\ell=2j_\ell + 2\gamma \mod 2k_\ell$, for $\ell=1,2$, see eq.~(\ref{dimchiralMM}).

For every asymmetric Gepner model, we first list the elements of the left (anti)chiral ring that satisfy the extra conditions~(\ref{asanti2}) in some twisted sectors $\gamma$. For the states that remain, we look for the minimal right conformal dimensions with the given quantum numbers $[2j_1,2j_2,2j_3,2j_4]_\gamma$. 
There are {\it a priori} three possibilities:
\begin{enumerate}
\item The contribution of the WZW models $\bar{\Delta}_\textsc{w}=j_3(j_3+1)/k_3+ j_4(j_4+1)/k_4$ is already too large whatever 
the contribution of the first two minimal models is.
\item If there are candidates with $j_3=j_4=0$, the contribution of the WZW models vanishes hence we get a subset of the chiral rings of 
the original symmetric Gepner model.
\item If there exists states with $0<\bar{\Delta}_\textsc{w}\leqslant 1/2$ one needs to check explicitly the overall smallest 
conformal dimension with the given quantum numbers.
\end{enumerate}

\subsection{Massless spectra for all asymmetric Gepner models}
Setting aside for convenience the two Gepner models constructed with six minimal models --~that are actually free theories so 
can be studied using free-fermion constructions~-- we have to consider, for each of the 14 remaining 
models, all inequivalent ways of choosing the two minimal models that are 'twisted' in our construction. Overall, one 
gets a list of 62 models which are given in table~\ref{modelslist}. In our conventions, a model $(k_1,k_2,k_3,k_4)$ has 
its last two minimal models, at levels $k_3$ and $k_4$, asymmetrized. 
\begin{table}[!ht]
\begin{center}
(2, 3, 10, 15), (2, 10, 3, 15), (2, 15, 3, 10), (3, 10, 2, 15), (3, 15, 2, 10), (10, 15, 2, 3)\\
(2, 3, 8, 24), (2, 8, 3, 24), (2, 24, 3, 8), (3, 8, 2, 24), (3, 24, 2, 8), (8, 24, 2, 3)\\
(2, 3, 9, 18), (2, 9, 3, 18), (2, 18, 3, 9), (3, 9, 2, 18), (3, 18, 2, 9), (9, 18, 2, 3)\\
(2, 3, 7, 42), (2, 7, 3, 42), (2, 42, 3, 7), (3, 42, 2, 7), (7, 42, 2, 3), (3, 7, 2, 42)\\
(2, 4, 6, 12), (2, 6, 4, 12), (2, 12, 4, 6), (4, 12, 2, 6), (6, 12, 2, 4), (4, 6, 2, 12)\\
(2, 4,5, 20), (2, 5, 4, 20), (2, 20, 4,5), (5, 20, 2, 4), (4, 5, 2, 20), (4, 20, 2, 5)\\
(2, 3, 12, 12), (2, 12, 3, 12), (12, 12, 2, 3), (3, 12, 2,12)\\
(3, 3, 4, 12), (3, 4, 3, 12), (3, 12, 3, 4), (4, 12, 3, 3)\\
(2, 5, 5, 10), (2, 10, 5, 5), (5, 5, 2, 10), (5, 10, 2, 5)\\
(2, 4, 8, 8), (2, 8, 4, 8), (8, 8, 2, 4), (4, 8, 2, 8)\\
(3, 4, 4, 6), (3, 6, 4, 4), (4, 4, 3, 6), (4, 6, 3, 4)\\
(3, 3, 6, 6), (3, 6, 3, 6), (6, 6, 3, 3)\\
(2, 6, 6, 6), (6, 6, 2, 6)\\
(4, 4, 4, 4)
\end{center}
\caption{\it List of all inequivalent asymmetric K3 Gepner models, the last two minimal models being the asymmetric ones 
in each case.}
\label{modelslist}
\end{table}

\subsection*{STU models}

Among those listed in table~\ref{modelslist}, one finds that 33 models are actually free of any massless modulus from the 
asymmetrized $K3$ Gepner model; these are the models that do not appear in appendix~\ref{appspectra}. 

In all cases, the $T$ and $U$ moduli of the two-torus and the axio-dilaton modulus $S$ are still part of the spectrum; they 
are obtained by taking the identity operator in the asymmetric Gepner model, and vanishing left and right momenta along the twisted 
$T^2$. The massless spectrum contains also four $U(1)$ gauge fields coming from the Kaluza-Klein reduction on the two-torus 
(being states with zero momentum along the $T^2$ they also survive the twist). In order to organize these degrees of 
freedom and their fermionic partners in $\mathcal{N}=2$ multiplets, we have no choice but to consider that the three complex 
scalars $S$, $T$ and $U$ are part of vector multiplets. Hence unlike for type II compactifications on Calabi-Yau three-folds 
the dilaton belongs to a vector multiplet. 

Therefore these models flow at low energies to the so-called $STU$ model of $\mathcal{N}=2$ supergravity~\cite{Cremmer:1984hj,Duff:1995sm}, 
which contains besides the supergravity multiplet only three Abelian vector multiplets, whose scalar components are denoted 
$S$, $T$ and $U$.

\subsection*{Models with surviving moduli}

In the other 29 models, that are given in appendix~\ref{appspectra}, some massless hypermultiplets remain in the spectrum. 
The massless states of any  such asymmetric Gepner model are given by the subset of massless states in the associated 
symmetric Gepner model that satisfy the extra conditions~(\ref{asanti2}). 

As was argued before it may have been possible to find some massless states that do not belong to the right (anti-)chiral ring 
in the asymmetric models, as the right R-charge is not integer-valued. The analysis done for all models, whose results 
are summarized in app.~\ref{appspectra}, shows that there aren't any other massless states besides the truncated 
chiral rings. There is no particular relation between the right chiral and antichiral rings as the right $\mathcal{N}=4$ 
superconformal symmetry of the Gepner model factor is broken. We indeed find that the dimension of these two rings do not 
match generically.

Interestingly, since the dilaton belongs to a vector multiplet, the hypermultiplet moduli space receives no quantum corrections, 
unlike in CY$_3$ compactifications. This is related to the underlying $\mathcal{N}=4$ supergravity theory corresponding 
to the symmetric Gepner model at low energies, even though the gravitini masses, set by the $T^2$ moduli, are not necessarily 
small hence the breaking not necessarily of the 'spontaneous' type.

\subsection{Moduli spaces}
To summarize, the moduli space of a given asymmetric Gepner model splits into two  subspaces. 
The first one (hypermultiplets moduli space) is spanned by the leftover moduli of $K3$ that are not lifted by the fibration, if 
there are any. As we have noticed before, all the $RR$ ground states 
are lifted, so there are no moduli coming from RR $p$-forms integrated over cycles of $K3$. The second subspace  
(vector multiplets moduli space) is spanned by the torus moduli $T$ and $U$, which 
survive in all models, and by the axio-dilaton $S$. 

The asymmetric models were built by choosing particular values for the torus moduli, using an orthogonal two-torus 
with $R_\textsc{x} = \sqrt{\alpha' k_3}$ and  $R_\textsc{y} = \sqrt{\alpha' k_4}$ and no B-field, 
see eq.~(\ref{charcombis}). As the $T$ and $U$ moduli are not lifted, it is possible to reach any value for 
them by exact marginal deformations built out of the $U(1)^2$ left- and right-moving currents. The masses of the 
lifted $K3$ moduli and of the massive gravitini are therefore generically functions 
of the complex and K\"ahler structure of the two-torus. In particular in the decompactification limit $U/\alpha'\to \infty$ one 
finds a $K3\times \mathbb{R}^2$ with all the moduli of the original $K3$ Gepner model restored. In general the masses of the two 
massive gravitini are given by 
\begin{equation}
M_{\psi_\mu} (T,U) = \sqrt{\frac{T_2}{U_2}+\frac{(T_1\pm 1)^2}{U_2 T_2}}\, .
\end{equation}

At the point in the moduli space where we originally defined the models, namely $U=i\alpha'  \sqrt{k_3 k_4}$ and $T=i\sqrt{k_4/k_3}$, 
there is a non-Abelian symmetry enhancement.  The two asymmetric minimal models used in the construction have indeed an 
unbroken affine right-moving $SU(2)_{k_3} \times SU(2)_{k_4}$ symmetry, as can be seen from 
equation~(\ref{charcomba}).\footnote{This enhancement is distinct from what happens while compactifying at the self-dual radius.} 
Moving away from these values by marginal deformations will naturally  break this $SU(2)^2$ 
symmetry.

In our scan of all asymmetric models, we have found that all space-time massless scalars transform in the trivial 
representation of both $SU(2)$'s. Hence this symmetry, while not visible in the supergravity limit, is an organizing principle for 
the massive states. Likewise there are no massless space-time gauge fields that would make this symmetry local.

\section{Conclusions and perspectives}
\label{sec:concl}
In this work we have constructed a large class of type II compactifications with $\mathcal{N}=2$ supersymmetry, using 
asymmetric Gepner models. Interestingly, the space-time supercharges come only from the left-movers, indicating that these 
constructions should be somehow related to asymmetric freely-acting orbifolds of $K3\times T^2$ at Gepner points. 

More than half of the models have no massless hypermultiplets in their spectrum, reproducing at low energies 
$\mathcal{N}=2$ supergravity with $STU$ vector multiplets (having as scalar components the axio-dilaton and the torus 
moduli), and no other massless fields. The remaining models have a hypermultiplets moduli space 
which receives no quantum corrections as the dilaton sits in a vector multiplet, and whose dimension is model-dependent.

A very interesting generalization of this work would be to consider orientifolds of these models, for example 
analogues of the well-studied type IIB $K3\times T^2$ compactifications with $O7$-planes (using the 
involution $\Omega (-)^{F_L} \, \mathcal{I}_{T^2}$ where $\mathcal{I}_{T^2}$ 
is the inversion along the two-torus), $D7$-branes, $D3$-branes and/or fluxes~\cite{Dasgupta:1999ss}. 
In the case of symmetric Gepner models, orientifolds have been constructed  $e.g.$ in~\cite{Brunner:2004zd}; similar 
techniques can be used here. In our asymmetric Gepner models space-time supersymmetry comes only from the left-movers, 
and there are no RR fluxes, hence one can wonder whether such models are actually supersymmetric. 
As a preliminary step, we have considered boundary states for 
D7-branes, generalizing the results of~\cite{Recknagel:1997sb}, and obtained that their open string spectrum 
is not supersymmetric. Instead of having a generalized GSO-projection like~(\ref{masscondit}), giving a spectrum of 
integer R-charges, one gets the weaker condition
\begin{equation} 
\sum_{i=1}^4 \frac{m_i}{k_i} + \frac{M_3}{k_3} + \frac{M_4}{k_4} \in \mathbb{Z} \, ,
\end{equation}
where $M_{3,4}$ are the $U(1)_{k_3} \times U(1)_{k_4}$ charges along the $T^2$. The presence of fractional R-charges 
pinpoints the absence of supersymmetry. It would be very interesting to study these orientifold compactifications in more detail, in particular to 
check whether the open string spectrum on branes contains tachyonic states. It is worthwhile to notice finally that D7-branes are 
not strictly necessary as there is no RR tadpole to cancel. Hence unoriented models without open strings are possible, 
the price to pay being that the closed vacuum is corrected by the dilaton tadpole. 

Another important problem is the understanding of non-perturbative dualities using these compactifications as starting points, 
more precisely to find dual descriptions under STU triality~\cite{Duff:1995sm}. In particular, heterotic duals should be given by 
some twisted tori compactifications with nonperturbative duality twists (as $U$ and $S$ are exchanged). 

Finally, the (non-)geometric interpretation of the asymmetric Gepner models is also quite interesting to consider, 
if one is able to find suitable 'R-fluxes' and 'Q-fluxes' added to $K3\times T^2$ that could reproduce the 
spectra that we obtained, at the effective action level, using for instance the ten-dimensional formulation of 
non-geometric backgrounds developped in~\cite{Andriot:2011uh,Andriot:2013xca} (see in particular~\cite{Condeescu:2013yma} 
for a study of free-fermions models). In our case, having a clear understanding of the 
original symmetric Gepner model and of the potentially surviving moduli in terms of the $K3$ geometry 
would give some interesting insights on string (non-)geometry.

\section*{Acknowledgements}
We would like to thank Luca Carlevaro, Nick Halmagyi, Josh Lapan, Michela Petrini and Jan Troost for enlightening 
and entertaining discussions. This work was supported by French state funds managed by the ANR within the Investissements 
d'Avenir programme under reference ANR-11-IDEX-0004-02.

\appendix

\section{Massless moduli for all models}
\label{appspectra}
We provide below the complete list of massless moduli in all the models that actually admit massless 
scalars in their spectra besides $S$, $T$ and $U$, among those given in table~\ref{modelslist}.  All others models 
are free of hypermultiplets. 

As in the text we use the notation $(k_1,k_2,k_3,k_4)$ for the models themselves, the last two minimal models 
being the asymmetric ones, and $[2j_1,2j_2,2j_3,2j_4]_{\gamma}$ to label the massless states by their $SU(2)$ spins and 
twisted sector $\gamma$. In all cases one finds that massless states in the untwisted sector 
($\gamma=0$) belong to the $(a,a)$ and $(c,a)$ rings while states in the twisted sectors 
($\gamma\neq 0$) belong to the $(a,c)$ and $(c,c)$ rings. In all models, we did not find any non-chiral massless state. 

\subsection*{(2,3,10,15) family}
\begin{itemize}
\item (3,15,2,10)~: $[1, 10, 0, 0]_0$ and $[1, 10, 0, 0]_{20}$ 
\item (10,15,2,3)~: $[2,12,0,0]_0$, $[2,12,0,0]_{18}$,
$[4, 9, 0, 0]_0$, $[4, 9, 0, 0]_6$, 
$[6, 6, 0, 0]_0$, $[6, 6, 0, 0]_{24}$, 
$[8, 3, 0, 0]_0$ and $[8, 3, 0, 0]_{12}$
\end{itemize}

\subsection*{(2,3,8,24) family}
\begin{itemize}
\item (3, 24, 2, 8)~: $[1, 16, 0, 0]_0$ and   $[1, 16, 0, 0]_8$
\item (8, 24, 2, 3)~: $[1, 21, 0, 0]_0$ 
\end{itemize}

\subsection*{(2,3,9,18) family}
\begin{itemize}
\item (3, 9, 2, 18)~: $[1, 6, 0, 0]_0$
\item (3, 18, 2, 9)~: $[1, 12, 0, 0]_0$
\item (9, 18, 2, 3)~: $[1, 16, 0, 0]_0$, 
$[2, 14, 0, 0]_0$ and $[3, 12, 0, 0]_0$,
$[3, 12, 0, 0]_6$, $[4, 10, 0, 0]_0$, 
$[5,8,0,0]_0$, $[6, 6, 0, 0]_0$ and
$[6, 6, 0, 0]_{12}$ 
\end{itemize}

\subsection*{(2, 4, 6, 12) family}
\begin{itemize}
\item (4, 12, 2, 6)~: $[1, 9, 0, 0]_0$, 
$[2, 6, 0, 0]_0$ and $[2, 6, 0, 0]_6$
\item (6, 12, 2, 4)~: $[1, 10, 0, 0]_0$, 
$[2, 8, 0, 0]_0$,  $[2, 8, 0, 0]_4$, 
$[3, 6, 0, 0]_0$, 
$[4, 4, 0, 0]_0$ and  $[4, 4, 0, 0]_8$ 
\item (4, 6, 2, 12)~:  $[2, 3, 0, 0]_0$ 
\end{itemize}

\subsection*{(2, 4,5, 20) family}

\begin{itemize}
\item (5, 20, 2, 4)~: $[1, 16, 0, 0]_0$, $[1, 16, 0, 0]_4$, 
$[2, 12, 0, 0]_0$, $[2, 12, 0, 0]_8$, 
$[3, 8, 0, 0]_0$ and $[3, 8, 0, 0]_{12}$
\item (4, 20, 2, 5)~: $[1, 15, 0, 0]_0$, $[2, 10, 0, 0]_0$ and  $[2, 10, 0, 0]_{10}$  
\end{itemize}

\subsection*{(2, 3, 7, 42) family}

\begin{itemize}
\item (3, 42, 2, 7)~:  $[1, 28, 0, 0]_0$ and $[1, 28, 0, 0]_{14}$
\item (7, 42, 2, 3)~: $[1, 36, 0, 0]_0$, $[1, 36, 0, 0]_6$,
$[2, 30, 0, 0]_0$, $[2, 30, 0, 0]_{12}$,
$[3, 24, 0, 0]_0$, $[3, 24, 0, 0]_{18}$, 
$[4, 18, 0, 0]_0$, $[4, 18, 0, 0]_{24}$,
$[5, 12, 0, 0]_0$ and $[5, 12, 0, 0]_{30}$
\end{itemize}

\subsection*{(2, 3, 12, 12) family}
\begin{itemize}
\item (12, 12, 2, 3)~: $[2, 10, 0, 0]_0$  
$[3, 9, 0, 0]_0$,  
$[4, 8, 0, 0]_0$, 
$[5, 7, 0, 0]_0$, 
$[6, 6, 0, 0]_0$,
$[6, 6, 0, 0]_6$,
$[7, 5, 0, 0]_0$, 
$[8, 4, 0, 0]_0$,
$[8, 4, 0, 0]_6$,
$[9, 3, 0, 0]_0$ and 
$[10, 2, 0, 0]_0$
\item  (3, 12, 2,12)~: $[1, 8, 0, 0]_0$
\end{itemize}

\subsection*{(3, 3, 4, 12) family}
\begin{itemize}
\item (3, 12, 3, 4)~: $[1, 8, 0, 0]_0$
\item (4, 12, 3, 3)~: $[1, 9, 0, 0]_0$, $[1, 9, 0, 0]_3$
$[2, 6, 0, 0]_0$ and $[2, 6, 0, 0]_6$
\end{itemize}

\subsection*{(2, 5, 5, 10) family}
\begin{itemize}
\item (5, 5, 2, 10)~: $[2, 3, 0, 0]_0$ and 
$[3, 2, 0, 0]_0$ 
\item (5, 10, 2, 5)~: $[1, 8, 0, 0]_0$, $[2, 6, 0, 0]_0$ and $[3, 4, 0, 0]_0$ 
\end{itemize}

\subsection*{(2, 4, 8, 8) family}
\begin{itemize}
\item (8, 8, 2, 4)~: $[2, 6, 0, 0]_0$, $[3, 5, 0, 0]_0$, 
$[4, 4, 0, 0]_0$, $[4, 4, 0, 0]_4$, 
$[5, 3, 0, 0]_0$ and $[6, 2, 0, 0]_0$ 
\item (4, 8, 2, 8)~: $[1, 6, 0, 0]_0$ and $[2, 4, 0, 0]_0$
\end{itemize}

\subsection*{(3, 4, 4, 6) family}
\begin{itemize}
\item (3, 6, 4, 4)~: $[1, 4, 0, 0]_0$ and $[1, 4, 0, 0]_8$ 
\item (4, 4, 3, 6)~: $[2, 2, 0, 0]_0$ and $[2, 2, 0, 0]_6$  
\item (4, 6, 3, 4)~: $[2, 3, 0, 0]_0$ 
\end{itemize}

\subsection*{(3, 3, 6, 6) family}
\begin{itemize}
\item (3, 6, 3, 6)~: $[1, 4, 0, 0]_0$ 
\item (6, 6, 3, 3)~: $[2, 4, 0, 0]_0$ and   
$[3, 3, 0, 0]_0$,  $[3, 3, 0, 0]_3$,
and $[4, 2, 0, 0]_0$ 
\end{itemize}

\subsection*{(2, 6, 6, 6) family}
\begin{itemize}
\item (6, 6, 2, 6)~: $[2, 4, 0, 0]_0$, $[3, 3, 0, 0]_0$ and $[4, 2, 0, 0]_0$  
\end{itemize}

\subsection*{(4, 4, 4, 4) family}
\begin{itemize}
\item (4,4,4,4)~: $[2, 2, 0, 0]_0$ 
\end{itemize}

\section{$\boldsymbol{{\mathcal N}=2}$ characters}
\label{appchar}
The characters of the $\mathcal{N} =2$ minimal models with $c=3-6/k$, {\it i.e.}  
the supersymmetric $SU(2)_k / U(1)$ gauged \textsc{wzw} model, 
are conveniently defined through the characters $C^{j\ (s)}_{m}$ of the $[SU(2)_{k-2} \times U(1)_2] / U(1)_k$ bosonic 
coset, obtained by splitting the Ramond and Neveu--Schwarz 
sectors according to the fermion number mod 2~\cite{Gepner:1987qi}. Defining $q=e^{2\pi i\tau}$ and $z=e^{2\pi i\nu}$, 
these characters are determined implicitly through the
identity:
\begin{equation} \label{theta-su2}
\chi_{k-2}^{j} (\nu|\tau)
\Theta_{s,2}(\nu-\nu'|\tau) = \sum_{m \in \zi_{2k}} C^{j\ (s)}_{m} (\nu'|\tau)  \Theta_{m,k} \big(\nu-\tfrac{2\nu'}{k}\big|\tau\big) \, ,
\end{equation}
in terms of the theta functions of $\widehat{\mathfrak{su}(2)}_k$:
\begin{equation}\label{thSU2}
\Theta_{m,k} (\tau,\nu) = \sum_{n}
q^{k\left(n+\tfrac{m}{2k}\right)^2}
z^{k \left(n+\tfrac{m}{2k}\right)}\,,  \qquad m \in \mathbb{Z}_{2k} 
\end{equation}
and $\chi^j_{k-2} $ the characters of the affine algebra $\widehat{\mathfrak{su}(2)}_{k-2}$:
\begin{equation}\label{su2-char}
\chi_{k-2}^j (\nu|\tau) = \frac{\Theta_{2j+1,k} (\nu|\tau)-\Theta_{-(2j+1),k} (\nu|\tau)}{i\vartheta_1(\nu|\tau)}\,.
\end{equation}
Highest-weight representations are labeled by  $(j,m,s)$, corresponding to primaries of 
$SU(2)_{k-2}\times U(1)_k \times U(1)_2$. The following identifications apply:
\begin{equation}
\label{symMM}
(j,m,s) \sim (j,m+2k,s)\sim
 (j,m,s+4)\sim
 \big(\tfrac{k}{2}-j-1,m+k,s+2\big)
\end{equation}
as  the selection rule $2j+m+s =  0  \mod 2$. The spin $j$ is restricted to $0\leqslant j \leqslant \tfrac{k}{2}-1$.  
The conformal weights of the superconformal primary states are:
\begin{subequations}
\label{dimchiralMM}
\begin{align}
\Delta &=  \frac{j(j+1)}{k} - \frac{n^2}{4k} + \frac{s^2}{8}\ & \text{for} & \ -2j \leqslant n-s \leqslant 2j \\
\Delta &=  \frac{j(j+1)}{k} - \frac{n^2}{4k} + \frac{s^2}{8} + \frac{n-s-2j}{2}
\ & \text{for} & \  2j \leqslant n-s \leqslant 2k-2j-4
\end{align}
\end{subequations}
and their $R$-charge reads:
\begin{equation}
Q_R = \frac{s}{2}-\frac{m}{k} \mod 2 \,. 
\end{equation}

\paragraph{Chiral primary states} are obtained for $m=2(j+1)$ and 
$s=2$ (thus odd fermion number). Their conformal dimension reads:
\begin{equation}
\Delta= \frac{Q_R}{2} = \frac{1}{2} - \frac{j+1}{k}\, .
\end{equation}
\paragraph{Anti-chiral primary states} are obtained for $m=2j$ and $s=0$ 
(thus even fermion number). Their conformal dimension reads:
\begin{equation}
\Delta= -\frac{Q_R}{2} = \frac{j}{k}\, .
\end{equation}
The usual Ramond and Neveu--Schwarz characters are  obtained as:
\begin{equation}
C^{j}_{m} \oao{a}{b} (\nu|\tau)=  e^{\frac{i\pi ab}{2}} \left[ C^{j\, (a)}_{m} (\nu|\tau)
+(-)^b C^{j\, (a+2)}_{m}(\nu|\tau) \right],
\end{equation}
where $a=0$ (resp. $a=1$) denote the \textsc{ns} (resp. \textsc{r}) sector, and characters 
with $b=1$ are twisted by $(-)^F$. In terms of these characters one has the reflexion symmetry:
\begin{equation}
C^j_m \oao{a}{b}(\nu|\tau) = (-)^b C^{\tfrac{k}{2}-j-1}_{m+k} \oao{a}{b}(\nu|\tau)\, . 
\label{reflsym}
\end{equation} 
\subsection*{Modular transformations}
The $S$ and $T$ transformations give
\begin{subequations}
\allowdisplaybreaks
\begin{align}
C^j_m \oao{a}{b} (-1/\tau) &=  e^{\frac{i\pi}{2} ab} \frac{1}{\sqrt{2k}}
\sum_{n \in \mathbb{Z}_{2k}} e^{ \frac{i \pi mn}{k}} 
\sum_{j'} S^j_{\, j'} C^{j'}_{n} \oao{b}{-a} (\tau) \\
C^j_m \oao{a}{b} (\tau+1) &= e^{2i\pi (\frac{j(j+1)}{k}-\frac{m^2}{4k} - \frac{a(a-2)}{8})}
C^j_m \oao{a}{a+b-1} (\tau)
\end{align}
\end{subequations} 
with $S^j_{\, j'}=  \sqrt{2/k} \sin \pi\frac{(1+2j)(1+2j')}{k}$. 
Let us consider now the full partition function for type IIB on $K3\times T^2$, the $K3$ being a Gepner model. 
Under an $S$ transformation one gets 
\begin{multline}
Z = \frac{1}{\tau_2^2 \eta^4 \bar{\eta}^4} \frac{1}{K} \sum_{\gamma,\delta \in \mathbb{Z}_{K}} 
\frac{1}{2} \sum_{a,b=0}^{1} (-)^{a+b} \frac{1}{2} \sum_{\bar{a},\bar{b}=0}^{1} (-)^{\bar{a}+\bar{b} }
\frac{\vartheta^2 \oao{b}{-a}}{\eta^2} \frac{\bar{\vartheta}^2 \oao{\bar{b}}{-\bar{a}}}{\bar{\eta}^2}\\
\prod_{i=1}^{4} \sum_{j_i,n_i}
e^{i\pi \gamma(-2\gamma+a-\bar a) \frac{n_i}{k_i}}
\, \,_{(k_i-2)} S^{j_i}_{\, j_i \prime} C^{j_i \prime}_{n_i+b} \oao{b}{-a} 
\, \,_{(k_i-2)} S^{j_i}_{\, \tilde{\jmath}_i \prime} \bar{C}^{\tilde{\jmath}_i \prime}_{n_i+\bar{b}+2\delta} \oao{\bar{b}}{-\bar{a}} 
\end{multline}
Therefore it is invariant as is seen after the obvious redefinitions $(a',b')=(b,-a)$ and $(\gamma ',\delta ' )=(\delta,-\gamma)$, 
using that 
$\sum_{j} S^{j}_{\, j \prime} S^{j}_{\, \tilde{\jmath} \prime} = \delta_{j \prime,\tilde{\jmath} \prime}$.
Let us now consider a $T$ transformation. One gets
\begin{multline}
Z = \frac{1}{\tau_2^2 \eta^4 \bar{\eta}^4} \frac{1}{K} \sum_{\gamma,\delta \in \mathbb{Z}_{K}} 
\frac{1}{2} \sum_{a,b=0}^{1} (-)^{b} \frac{1}{2} \sum_{\bar{a},\bar{b}=0}^{1} (-)^{\bar{b} }
\frac{\vartheta^2 \oao{a}{a+b-1}}{\eta^2} \frac{\bar{\vartheta}^2 \oao{\bar{a}}{\bar{a}+\bar{b}-1}}{\bar{\eta}^2}\\
\prod_{i=1}^{4} \sum_{j_i,m_i} 
e^{i\pi (2(\delta+\gamma)-(b+a)+\bar b +\bar a) \frac{m_i}{k_i}} 
C^{j_i}_{m_i+a} \oao{a}{a+b-1} 
\bar{C}^{j_i}_{m_i+\bar{a}+2\gamma} \oao{\bar{a}}{\bar a +\bar{b}-1} 
\end{multline}
After redefining $\delta ' = \delta +\gamma$ and $b'=b+a-1$ it is also invariant under $T$.

\bibliography{bibgepner}

\end{document}